\documentclass[12pt,preprint]{aastex}
\usepackage{subfigure}
\usepackage{graphicx}
\usepackage{amssymb}

\begin{document}

\title{Probing the Low Mass Stellar End of the $\eta$~Chamaeleontis Cluster}

\author{Inseok Song\altaffilmark{1}, B. Zuckerman\altaffilmark{1}}
\affil{Department of Physics and Astronomy\\
University of California, Los Angeles\\
Los Angeles, CA 90095--1562, USA\\
\altaffiltext{1}{a member of the Center for Astrobiology at UCLA}}

\author{M. S. Bessell}
\affil{Research School of Astronomy and Astrophysics\\
Institute of Advanced Studies\\
The Australian National University, ACT 2611, Australia}

\begin{abstract}
We have identified three faint new members of the $\eta
$~Chamaeleontis cluster. Spectral types of the new members are
estimated to be $\sim $M5 based on their TiO band strengths and
broadband colors. With an age of $5-8$\,Myr for the cluster, masses
of these new members are estimated to be $\sim0.08$\,M$_\odot$.
All three display strong Li~6708~\AA\  absorption and H$\alpha $
emission features including one with H$\alpha $ emission
equivalent width $\sim 60$~\AA\  along with HeI~6678 \& 7605~\AA\
emission features that are characteristics of classical T~Tauri
stars. 
\end{abstract}


\section{Introduction}

Recently, a compact open cluster was found only 97~pc from Earth
deep in the southern hemisphere \citep{Discovery, Mamajek2000}.
Compared to other known stellar groups, $\eta $~Chamaeleontis is
quite unique in many aspects.  Within $\sim 100$~pc of Earth,
stellar groups with a similar number of constituents (i.e., few
tens of members) ranging in age from $\sim8$ to $\sim300$\,Myr
(TW~Hydrae Association, the $\beta $~Pictoris moving group, the
Tucana/Horologium Association, and Ursa~Majoris group) are much
more diffuse than $\eta$~Cha.  The $\eta $~Cha cluster of 15 known
members \citep{Discovery, ECHA13-14} is compact (diameter$\lesssim
2$~pc) and very young (age $5-8$~Myr) which are great advantages
in thorough identification of its members.  Despite a sensitive
search for members in the central $40'$ diameter of the cluster
(ROSAT High Resolution Imager field of view, see
\citealt{Discovery}), currently known cluster membership is
heavily weighted toward high mass stars \citep{Discovery,
Mamajek2000, ECHA13-14}.  This skewed mass function may be
interpreted either as dynamical scattering of low mass stars to
the outskirts of the cluster or suppressed low mass star
formation.

Therefore, we initiated a photometric/spectroscopic search for
``missing'' low mass members of $\eta $~Cha over a wide area
(diameter$\sim 3.5$~pc) using 2MASS and USNO-B1 \citep{USNO-B1}
all-sky survey data.

\section{Target selection and Observation}

\begin{figure}
\begin{center}\includegraphics[width=0.95\columnwidth]{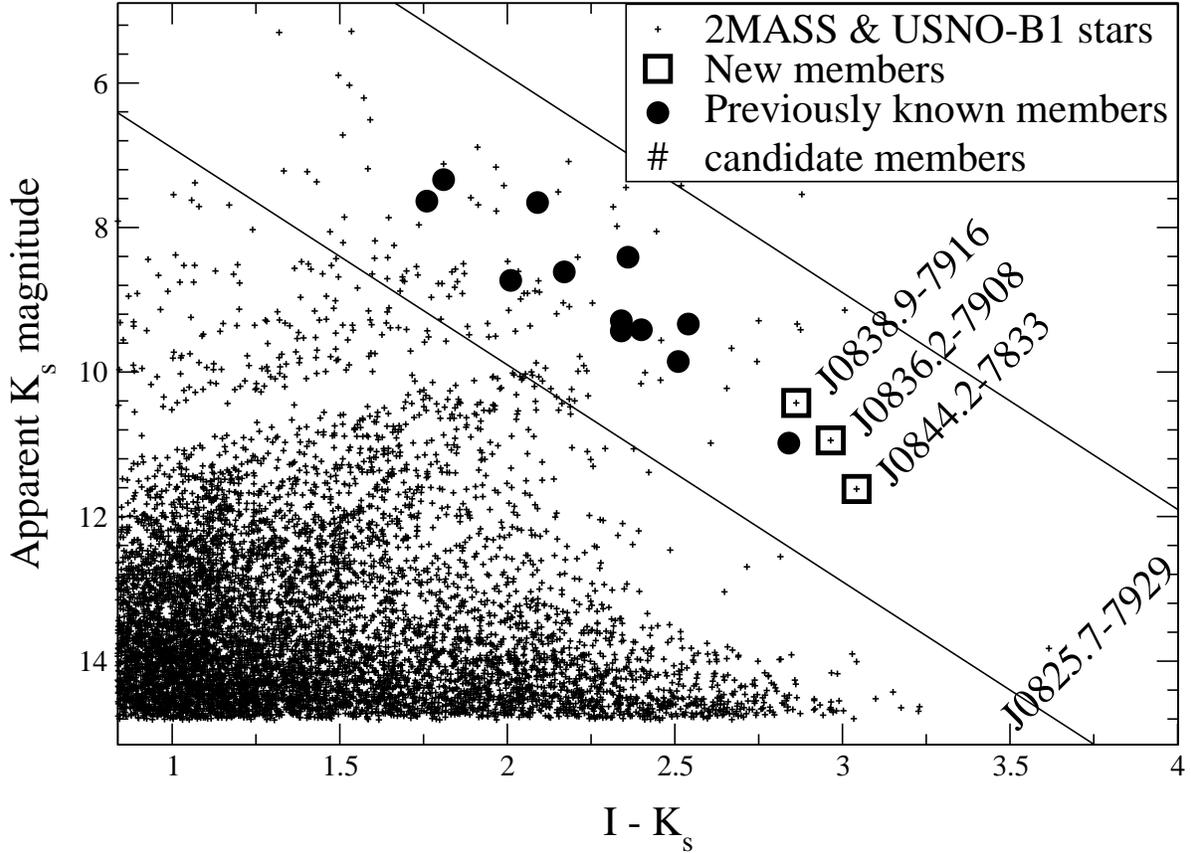}\end{center}
\caption{Color magnitude diagram of stars within a degree radius
of $\eta $~Cha.  $I$ and $K_{s}$ apparent magnitudes are from
USNO-B1 and 2MASS catalogs, respectively. Large solid circles
indicate already known cluster members and squares denote newly
identified members. Four plausible later spectral type cluster
member candidates are annotated.  To better define the location of
the cluster isochrone, we have used $I$-band magnitudes from
\cite{ECHA_photom} for the previously known members.  If USNO-B1
$I$ magnitudes are used instead, then the scatter among known
members is about $2-3$ times larger than shown here (some scatter
among known members is due to binaries which are not corrected
for). Two solid lines define the search region of this study.
Only stars with 2MASS $K_{s}$ photometry quality of `A' are
plotted. Three early-type previously known members have
$I-K_s<1.0$. A '$\times$' symbol indicates K/early-M candidate
members observed spectroscopically in addition to the four
late-type candidate members mentioned earlier.\label{CMD}}
\end{figure}

Since $JHK_{s}$ colors do not effectively differentiate among
stars of early to mid-M spectral types, to identify potential
members of the cluster, we first constructed an $I-K_{s}$ versus
$K_{s}$ color magnitude diagram using 2MASS (for $K_{s}$
magnitude) and USNO-B1 (for $I$ magnitudes) catalog data
(Figure~\ref{CMD}). Because of the small proper motions of cluster
members ($\sim 40$~mas/yr) and the small baseline between the
2MASS images and USNO-B1 plate epochs ($<20$\, years), we matched
2MASS and USNO-B1 sources only when offsets between the catalog
positions were $<1.5''$. The cross correlation was performed
within one degree radius of $\eta $~Cha (J2000, $08:41:19.51$,
$-78:57:48.1$).  USNO photographic data have large photometric
uncertainty of up to 0.5~mag, but because of the long baseline
afforded by the $I-K_{s}$ color, one can use this color reliably
even with occasional color uncertainty of up to 0.5~mag. 

All previously known cluster members are well separated from the
vast majority of background stars especially in the M spectral
type region ($I-K_{s}\gtrsim 2.0$). This is because of the
proximity and extreme youth of the cluster. Almost all background
stars appear fainter than the cluster members as one would expect
(Figure~\ref{CMD}). For clarity and to better define the locus of the
cluster members, we used I band data from \citet{ECHA_photom}
instead of the USNO-B1 I magnitudes for the already known members. 

The apparent avoided region ($\sim 0.5$~mag width in $K_{s}$)
around $K_{s}\sim 10.5$ of Figure~1 is due to an incorrect
photometric calibration of USNO-B1 catalog $I$-band magnitudes.
Photographic plates covering the $\eta$~Cha region are
``uncalibratable"; their intrinsic variation is many magnitudes
and their non-linearity is terrible even by photographic standards
(D. Monet 2003, private communication). When compared to DENIS~I
measurements for common stars in the region, we found that bright
stars (above the apparent gap) are systematically brighter by
$\sim$0.25\,mag whereas fainter stars are systematically fainter
by $\sim$0.25\,mag than their DENIS~I magnitudes. However, this
systematic error does not affect our study because our search
region on Figure~1 was chosen sufficiently large (4\,mag width in
$K_s$ for a given $I-K_s$ color) so that all cluster members whose
I measurements suffered by this systematic error should still be
selected.

Occasionally, foreground dwarfs and background giants appear near
the cluster members and we could identify most such bright
contaminants from their spectral types (from SIMBAD), Hipparcos
distance, proper motions, and brightness. 

In this study we focused mainly on late-type members of the
cluster for the following reasons. First, they are better
distinguished from non-members due to the paucity of such red,
bright contaminants (see Figure~\ref{CMD}). Secondly, if dynamical
segregation took place in the cluster, low mass members would be
expected to be found on the outskirts of the cluster.  Four
candidates that lie near the extrapolated cluster isochrone are
selected (Figure~\ref{CMD}).

To check the reliability of selecting candidate members based on
location on the color magnitude diagram, we took a spectrum of
candidate cluster member J$0838.9-7916$ with the Low-Dispersion Survey
Spectrograph (LDSS-2) at the Magellan Clay 6.5~m telescope at
Las~Campanas Observatory on UT date 2003-03-28.  Although the
spectral resolution of LDSS-2 was too low to check for the
presence of a Li~6708~\AA\ absorption feature, the spectrum showed
prominent Hydrogen Balmer series emission lines which indicated
likely cluster membership for this star. Encouraged by the
J$0838.9-7916$ observation, we observed all four candidate members
with the Royal Greenwich Observatory (RGO) Spectrograph and 1200R
grating at the AAT 3.9~m telescope on UT date 2003-05-10. RGO
spectra were reduced following standard procedure (bias
subtraction, flat fielding, etc.) using IRAF tasks.  Telluric and
instrumental response features were removed by using a spectrum of
smooth spectrum standard star, HD~84903. 

To check if there exist any previously unidentified K/early-M type
($1.0\lesssim I-K_s\lesssim2.2$) cluster members, we selected 15
target stars within a degree radius of $\eta$~Cha that meet the
following criteria: (1)\,locations on CMD lie between
$\pm2.0$\,mag of the locus of known members, (2)\,proper motions
from Tycho-2 or USNO-B1 similar to those of known cluster members
($-56<\mu_{RA}<-6$ and $4<\mu_{DEC}<54$), (3)\,I magnitudes from
USNO-B1 smaller than 13.0 to ensure high S/N data from the SSO
2.3\,m telescope. (4)\,locations on a $I-J$ versus $I-K_s$
color-color diagram are within $\pm0.5$\,mag of the locus of known
members.  These selected stars were observed during UT date
2003-07-11 through 07-13 with the double beam grating spectrograph
(DBS) of the Australian National University's 2.3\,m telescope.
DBS data were reduced in a similar fashion as described in
\citet{LDB}.  Positions and USNO-I magnitudes of these 15 stars are
given in Table~\ref{noMembers}.

\section{Results and Discussion}

\begin{table}

\caption{Four late-type candidate members of the $\eta $~Chamaeleontis cluster \label{table}}

\begin{tabular}{ccccc}
\hline 
Candidate No.&
\multicolumn{1}{c}{J$0838.9-7916$}&
\multicolumn{1}{c}{J$0836.2-7908$}&
\multicolumn{1}{c}{J$0844.2-7833$}&
\multicolumn{1}{c}{J$0825.7-7929$}\\
\hline
RA (J2000)&
$08:38:51.10$&
$08:36:10.72$&
$08:44:09.14$&
$08:25:41.65$\\
DEC (J2000)&
$-79:16:13.6$&
$-79:08:18.3$&
$-78:33:45.7$&
$-79:29:11.9$\\
EW(H$\alpha $) {[}\AA{]}&
$-10.3$&
$-7.6$&
$-57.8$&
absorption\\
EW(Li 6708) {[}m\AA{]}&
500&
470&
480&
---\\
TiO5&
0.29&
0.27&
0.33&
--\\
$I-K_{s}$&
2.86&
2.96&
3.04&
3.62\\
Spectral Type&
M5&
M5.5&
M4.5/M5.5&
--\\
Membership&
Likely&
Likely&
Likely&
No\\
\hline
\multicolumn{5}{l}{J$0844.2-7833$ also shows HeI $\lambda$6678
(EW=$-$1.8~\AA) and $\lambda$7605 (EW=$-$0.9~\AA).} \\
\multicolumn{5}{l}{Negative EW indicates an emission line}
\end{tabular}
\end{table}

\begin{figure*}
\begin{center}\includegraphics[  width=0.75\paperwidth]{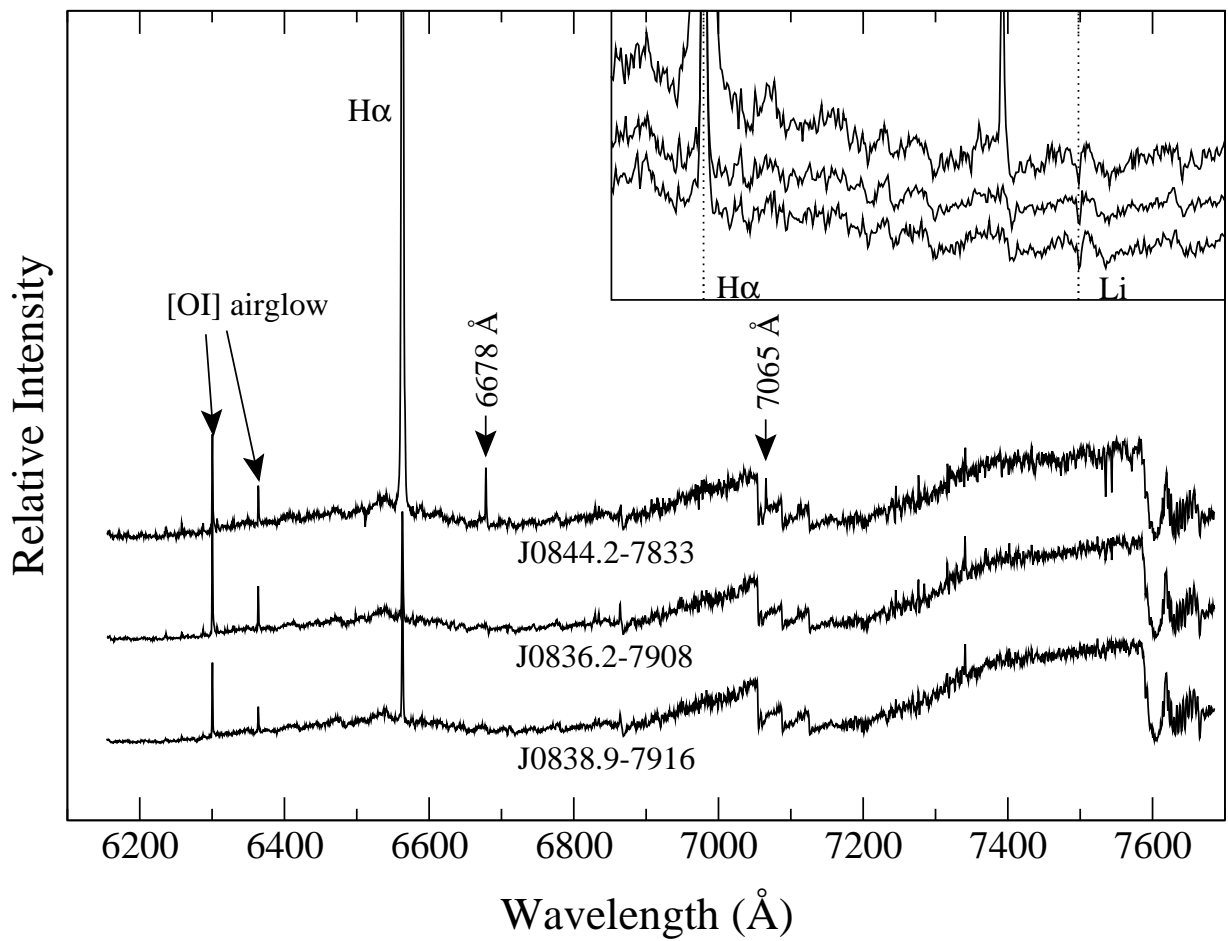}\end{center}
\caption{AAT RGO spectra of three newly identified members of the
$\eta $~Cha cluster. From top to bottom, candidates
J$0844.2-7833$, J$0836.2-7908$, \& J$0838.9-7916$ are plotted
(same as in the inset where lithium $6708$~\AA\ is indicated).
Emission lines at $\lambda 6300$ \& $6363$~\AA\ are air-glow. In
the spectrum of J$0844.2-7833$, HeI~$6678$ \& $7065$~\AA\ appear
in emission.\label{spectrum}}
\end{figure*}

RGO spectra of candidate members J$0838.9-7916$,
J$0836.2-7908$, \&
J$0844.2-7833$ all showed
prominent H$\alpha $ emission features and strong Li~6708~\AA\
line absorption (Figure~\ref{spectrum}) which implies true
membership. The two leftmost emission features (6300 \& 6363~\AA)
are Oxygen~I air-glows due to insufficient telluric features
removal (confirmed by looking at the sky region of the raw
spectra). The proper motion of J$0838.9-7916$ from
USNO-B1 catalog ($\mu _{RA}=-34$, $\mu _{DEC}=+24$~mas/yr) also
supports its true membership when compared to the mean proper
motions of known cluster members ($\mu _{RA}=-30$, $\mu
_{DEC}=+20$~mas/yr). Based on the TiO5 $[\equiv F(7126-7135\, \AA
)/F(7042-7046\, \AA )]$ band strength around 7100~\AA\
\citep{TiO5} and 2MASS $JHK_{s}$ measurements, we estimate
spectral types $\sim $M5 of the three stars marking these stars as
the lowest mass ($\sim0.08\,M_\odot$ based on $5-8$\,Myr pre-main
sequence model of \citealt{BCAH}) known members of the cluster.
Details are summarized in Table~\ref{table}.  

J$0825.7-7929$ turned out to be a close visual binary with $5''$
separation and we took spectra of both stars by putting them on
the slit simultaneously. The fainter star was a late-type star and
the brighter one was a background hot star. For the fainter star
of the J$0825.7-7929$ visual binary, even with the accumulated
integration time of one hour, the signal-to-noise ratio of the
reduced spectrum was too low ($<5$) to draw any firm conclusions.
However, from the raw data, we see a hint of an H$\alpha$ line in
absorption. Were J$0825.7-7929$ a true member, its spectral type would be
M7/8 and we would expect to see a prominent H$\alpha$ emission
feature. Thus, we believe the candidate J$0825.7-7929$ is a non-member. 

Because of the 2MASS sensitivity limit $(K_{s}\sim14.3$), we would
not be able to find any later M or L-type members even if they
exist. A deep near infrared photometric measurement can allow one
to search for L-type cluster members. If found, late-M/early-L
members will be brown dwarfs and late L members will be high mass planets. 

\begin{figure}
\begin{center}\includegraphics[  width=0.75\paperwidth]{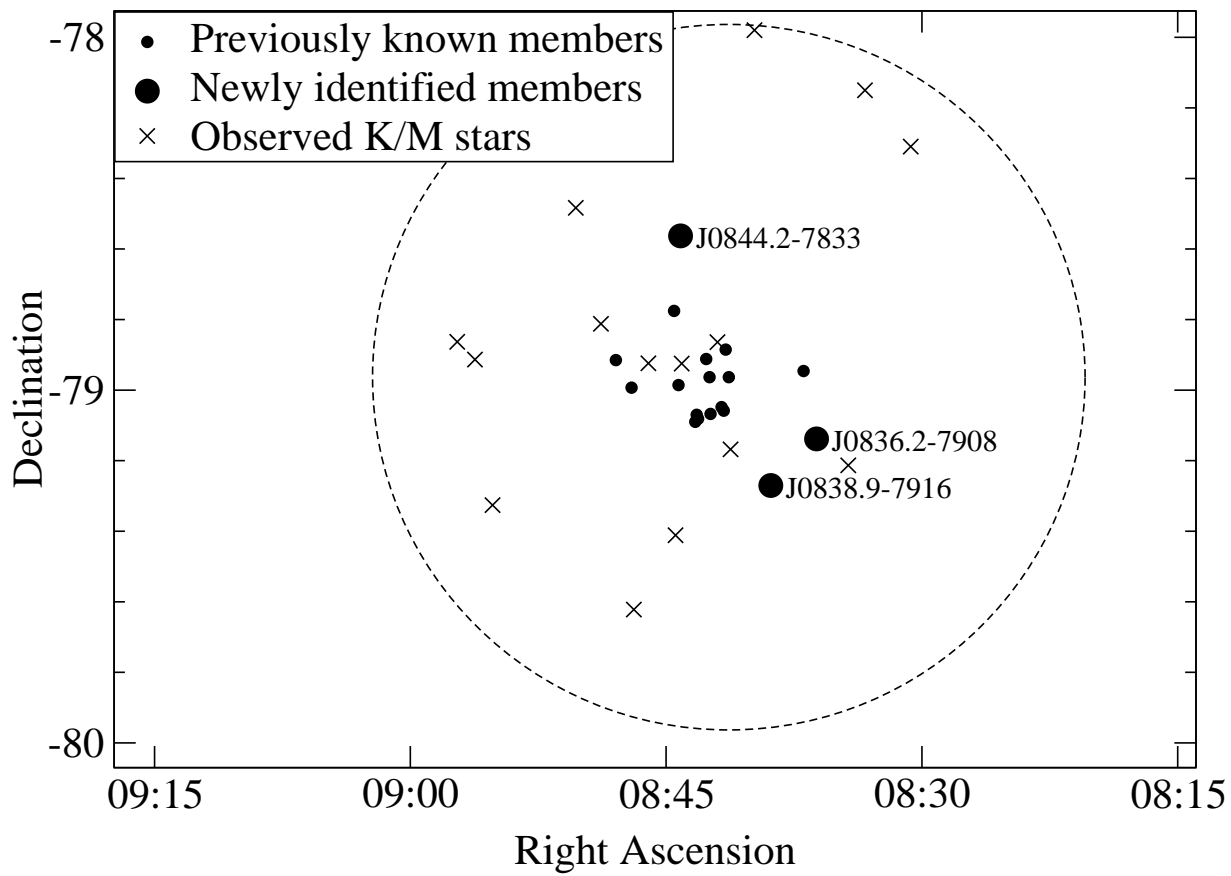}\end{center}
\caption{Positions and membership numbers of previously known
(small solid circles) and newly identified (large solid circles)
members of the $\eta $~Cha cluster. A '$\times$' symbol indicates
spectroscopically confirmed non-members with K/early-M $I-K_s$
colors as listed in Table~2. The dashed circle indicates a 1~degree radius circle
around $\eta $~Cha. \label{postion}}
\end{figure}

The fact that none of the 15 K/early-M candidate members turned
out to be a member is significant. Considering the selection
criteria of those 15 candidates, we expected at least a fair
fraction of the candidates to be real members. In addition to
these 15 stars, we observed seven stars satisfying the same set of
selection criteria but slightly outside of the one degree radius
search region. However, none of them turned out to be a member.
While three lowest mass members are found at the outskirt of the
cluster, the fact that no K/early-M members have yet been found in
the outer region of the cluster indicates that a strong mass
segregation may have occurred already in this very young cluster
(Figure~\ref{postion}).  A more thorough survey of cluster members
using better photometric data (i.e., full DENIS survey data;
currently available April 2003 released data cover 40~\% of the
1~degree radius region around $\eta $~Cha.) accompanied by
rigorous spectroscopic observation may help to test the mass
segregation of this cluster more thoroughly. Allegedly, JHK colors
from 2MASS survey do not effectively differentiate K/early-M stars
thus cluster candidate member selection based only on 2MASS data
is not useful.

\begin{table}
\caption{Spectroscopically confirmed non-members plotted on
Figure~3.  \label{noMembers}}
\begin{tabular}{ccc||ccc}
\hline
RA      &       DEC     & I     &RA     &DEC    &I      \\
\hline
%08:12:27.27 &-79:35:42.0 &  9.17&08:12:28.61 &-78:50:25.2 &  9.21 \\
%08:17:37.54 &-78:57:41.9 &  8.95&08:19:28.45 &-79:58:43.3 &  9.38 \\
08:30:39.17 &-78:18:35.0 & 12.60&08:33:20.24 &-78:09:00.1 &  9.37 \\ 
08:34:19.11 &-79:12:47.3 &  6.82&08:39:49.04 &-77:58:46.6 &  9.72 \\
08:41:13.27 &-79:10:04.4 &  9.13&08:41:59.27 &-78:51:50.4 & 12.56 \\
08:44:05.06 &-78:55:28.9 &  8.63&08:44:26.87 &-79:24:40.8 &  9.51 \\
08:46:02.37 &-78:55:26.2 & 10.97&08:46:53.58 &-79:37:19.1 & 10.27 \\
08:48:49.17 &-78:48:43.5 & 12.22&08:50:17.67 &-78:28:59.7 &  8.23 \\
08:54:23.64 &-79:46:28.4 &  9.09&08:55:10.63 &-79:19:33.7 &  8.93 \\
08:56:12.55 &-78:54:48.2 &  9.41 \\
%&08:57:14.87 &-78:51:47.2 & 10.78 \\
%09:08:29.91 &-78:28:51.5 & 10.35&09:08:48.01 &-78:49:37.6 &  8.65 \\
\hline
\end{tabular}
\end{table}

\begin{figure*}
\begin{center}\begin{tabular}{ccc}
\subfigure[J$0838.9-7916$]{\includegraphics[  width=0.25\paperwidth,
  keepaspectratio]{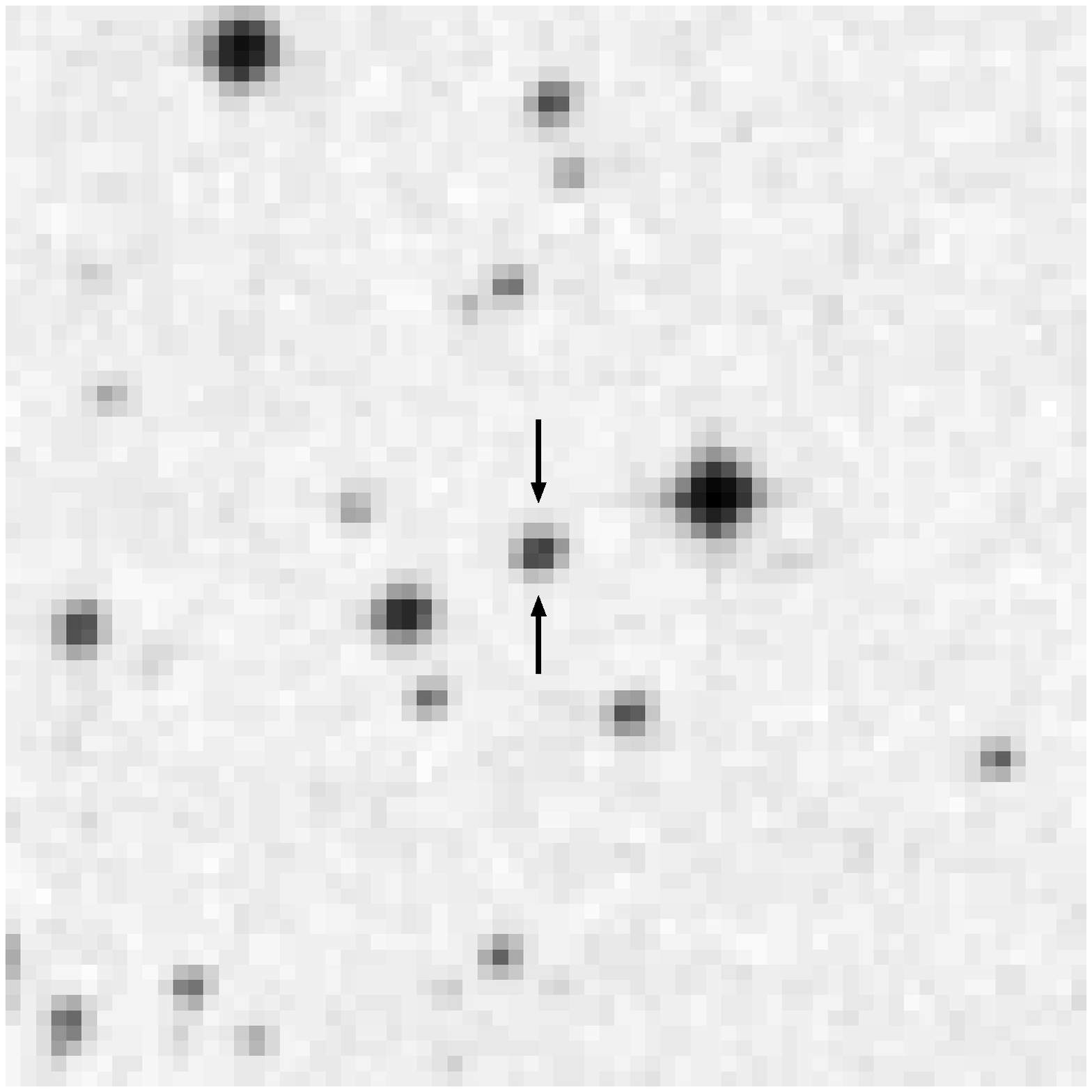}}&
\subfigure[J$0836.2-7908$]{\includegraphics[  width=0.25\paperwidth,
  keepaspectratio]{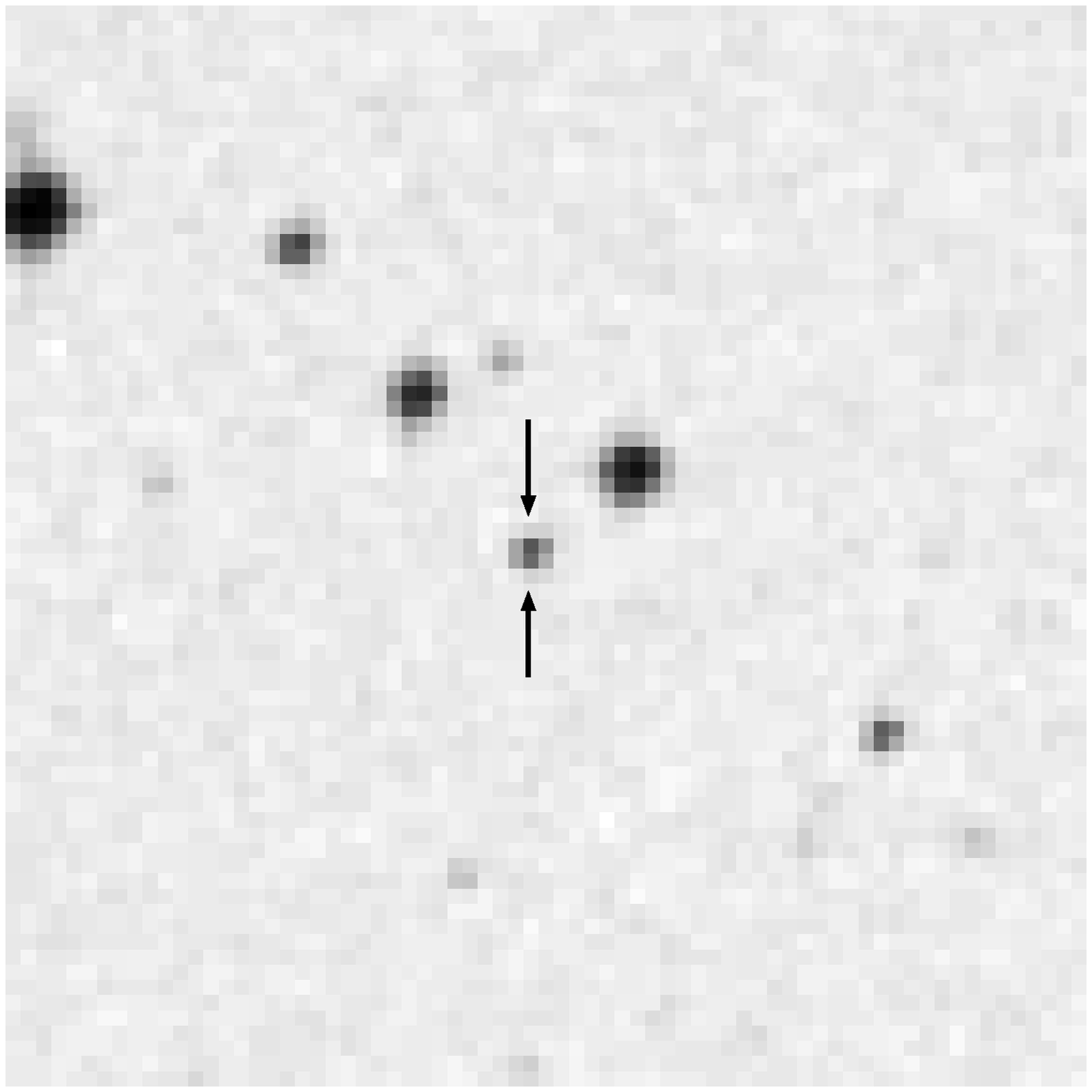}}&
\subfigure[J$0844.2-7833$]{\includegraphics[  width=0.25\paperwidth,
  keepaspectratio]{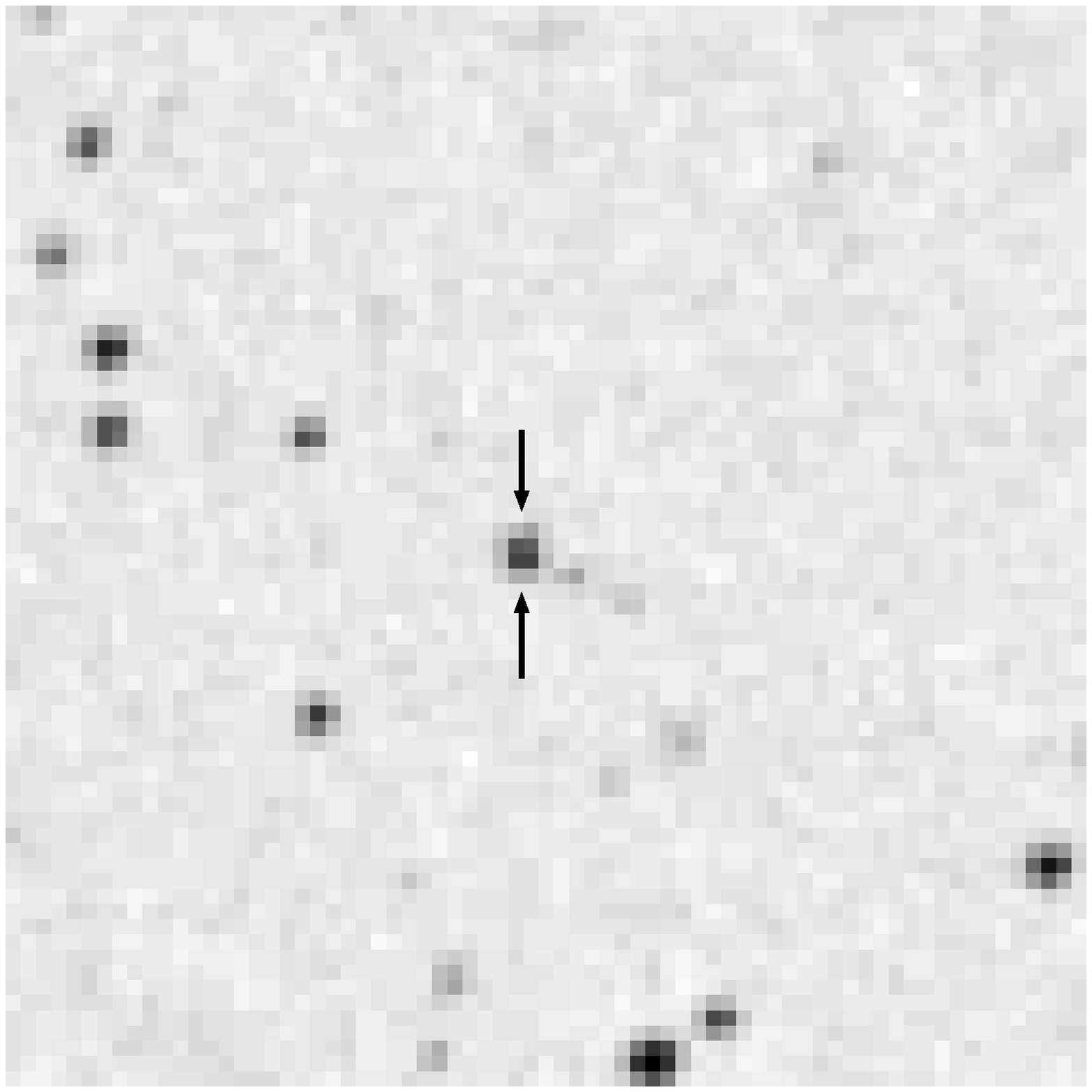}}\\
\end{tabular}\end{center}
\caption{Finder charts for three newly identified members of the $\eta $~Cha
cluster. Displayed are $2'\times 2'$ Digital Sky Survey images; north
up and east left.\label{finder}}
\end{figure*}

Finder charts of the three identified new members are provided in
Figure~\ref{finder}. 

Since high spectral resolution spectra are not available for $\eta
$~Cha members, the velocity dispersion of the cluster has not been
measured yet; hence we do not know if it is gravitationally bound
or not. However, compared to the more dispersed but similar age
stellar groups, the TW~Hyadrae association ($\sim 8$~Myr) and
$\beta $~Pictoris moving group ($\sim 12$~Myr), the $\eta $~Cha
cluster is quite compact suggesting a gravitationally bound
system. 

\acknowledgements{We thank Dr. Paul Martini for obtaining the
LDSS-2 spectra of J$0838.9-7916$.  We are also grateful to the director of
the Anglo-Australian Observatory, Dr. Brian Boyle, for rapidly
approving the use of Director's Discretionary Time and to Dr.
Raylee Stathakis for taking RGO spectra of the four candidates.
This research was supported in part by the UCLA Astrobiology
Institute and by a NASA grant to UCLA.}

\clearpage
\bibliographystyle{apjl}

\end{document}